\documentclass[preprint,prb,amsmath,amssymb,floatfix,draft]{revtex4}

\newcommand{\be}{\begin{equation}}
\newcommand{\ee}{\end{equation}}
\newcommand{\bn}{\begin{displaymath}}
\newcommand{\en}{\end{displaymath}}
\newcommand{\bs}{\begin{eqnarray}}
\newcommand{\es}{\end{eqnarray}}
\newcommand{\nn}{\nonumber}

\newcommand{\sn}[1]{\mbox{sn($#1$)}}
\newcommand{\cn}[1]{\mbox{cn($#1$)}}
\newcommand{\dn}[1]{\mbox{dn($#1$)}}
\newcommand{\di}{\Diamond}

\usepackage[final]{graphicx}

\begin{document}

\title{Geometrical aspects of the Z-invariant Ising model}

\author{Ruben Costa-Santos
\footnote{e-mail R.A.Costa-Santos@phys.uu.nl}}
\affiliation{Spinoza Institute, Utrecht University, Leuvenlaan 4, 3584 CE Utrecht}

\preprint{SPIN 03-40 ITPUU 03-63 }

\begin{abstract}

We discuss a geometrical interpretation of the  Z-invariant Ising model in terms of isoradial embeddings of planar lattices. The  Z-invariant Ising model can be defined on an arbitrary planar lattice if and only if certain paths on the lattice edges  do not intersect each other more than once or self-intersect. This topological constraint is equivalent to the existence of isoradial embeddings of the lattice. Such  embeddings are characterized by angles which can be related to the model coupling constants in the spirit of Baxter's geometrical solution. The Ising model on isoradial embeddings  studied recently by several authors in the context of discrete holomorphy corresponds to the critical point of this particular Z-invariant Ising model.

\end{abstract}

\maketitle

\section{Introduction}\label{section1}

The most general setting in which the two dimensional Ising model is exactly solvable is obtained when the model is invariant under star-triangle  deformations of the lattice structure. This model was introduced by Baxter \cite{baxter1,baxter2} and is known in the literature as the Z-invariant Ising model. 

Recently the Ising model on a certain class of planar lattices, those with isoradial embeddings as defined bellow, has attracted attention in connection with the concept of discrete holomorphy \cite{mercat,duffin,kenyon1}. Mercat \cite{mercat} conjectured a criticality condition and Kenyon \cite{kenyon3} showed that the restricted model is solvable. In this paper we will prove the criticality conjecture and bring these results into perspective by showing that the Ising model considered by these authors is the restriction to the critical point of a Z-invariant Ising model. We will show that isoradial embeddings provide a geometrical interpretation of the Z-invariant Ising model   which generalizes the model's geometrical solution \cite{baxter2} in terms of angles characterizing straight rapidity lines.

\begin{figure}[t]
\includegraphics[width=11cm]{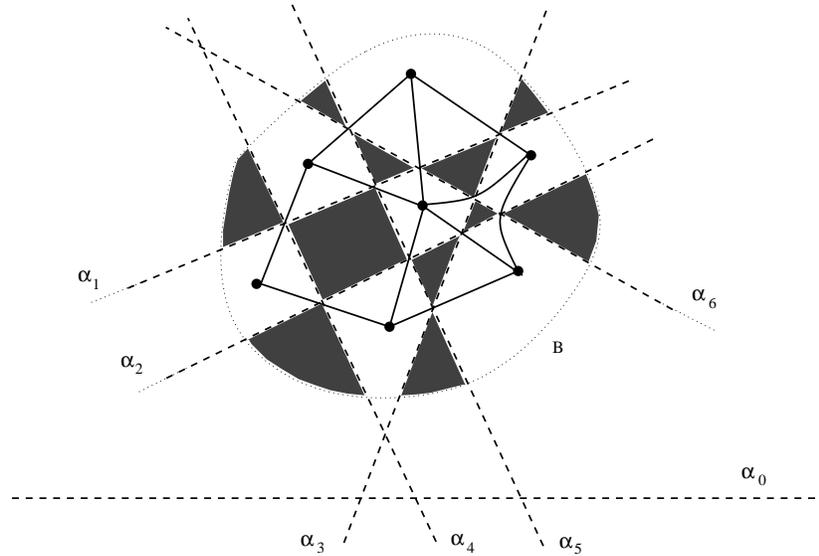}
\caption{An example of Baxter's  straight line lattice ${\cal G}_s$, shown in dashed line,  with each line having an associated rapidity $\alpha_i$. A Z-invariant Ising model can be defined on the resulting  lattice $G$, shown in full lines.}
\label{fig1}
\end{figure}

In the seminal work of Baxter \cite{baxter2} the Z-invariant Ising model is defined on planar graphs or `lattices'  $G$, which are defined using an associated rapidity lattice ${\cal G}_s$ of planar intersecting straight lines such that no three lines intersect at a point. These lattices ${\cal G}_s$ can be colored with two colors in such a way that faces with different colors do not share edges. The Z-invariant Ising model is defined on the lattice $G$  with  vertices on one of the colored subset of faces and edges connecting neighboring same color faces which share a vertex, for an example see of Fig. \ref{fig1}. The model coupling constants are determined by parameters, called the rapidities, associated with the straight lines in ${\cal G}_s$.
  
\begin{figure}[t]
\includegraphics[width=9cm]{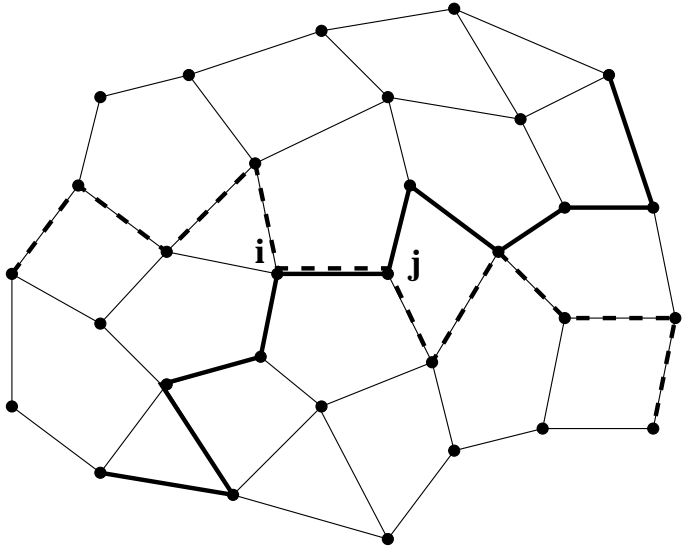}
\caption{An arbitrary planar lattice $G$ and the two zig-zag paths which intersect at the edge $[ij]$. Zig-zag paths on a planar lattice turn  alternately maximally right and left and are said to intersect when share a common edge.} 
\label{fig2}
\end{figure}

In this paper we will address the converse problem. Namely, given an arbitrary planar lattice $G$, defined without reference to a rapidity lattice, is it possible to assign  coupling constants to its edges in such a away that the Ising model on $G$ is Z-invariant? 
This problem has been considered previously in the context of quasi-crystals \cite{korepin1,korepin2}, by using either the de Bruijn grid method \cite{ned} or problem specific rapidity line choices \cite{baake}. For  general planar lattices we will show in Section \ref{section3} that a Z-invariant Ising model  can be defined if and only if the lattice satisfies a certain topological property: that zig-zag paths on the lattice edges,  this is paths which alternately turn maximally left and right (see Fig. \ref{fig2}), do not intersect each other more than once or self-intersect. If this condition is satisfied a Z-invariant Ising model can be defined using  rapidity lines in one to one correspondence with the zig-zag paths.

Zig-zag paths are closely related to  de Bruijn's \cite{ned} concept of skeleton of a parallelogram tilling and have recently been used by Kenyon and Schlenker\cite{kenyon2} to discuss the existence of isoradial embeddings of planar lattices. A strictly convex {\bf isoradial embedding} of a planar lattice is a particularly symmetric drawing of the lattice such that every vertex  is at unit distance from the center of the faces to which it belongs.  For instance  Fig. \ref{fig3} would be an isoradial embedding of the lattice shown if drawn in such a way that all edges in dotted line have equal unit length.
These embeddings are particularly useful in the context of discrete holomorphy \cite{duffin,mercat,kenyon1}, where the isoradiality condition  allows the definition of discrete analytic functions.

The intersection properties that we need to impose on the zig-zag paths of a planar lattice, to be able to define a Z-invariant Ising model,  are  equivalent \cite{kenyon2} to the existence of isoradial embeddings of that lattice. Therefore a  Z-invariant Ising model can defined on a planar lattice $G$ if and only if $G$ has isoradial embeddings.

This result can better understood by considering a geometric interpretation of the model. In Section \ref{section5} we will show that a Z-invariant Ising model, with non-vanishing ferromagnetic couplings, on a planar lattice defines an isoradial embedding of the lattice which is characterized by a set of geometric angles determined by the model's coupling constant at criticality. This geometric interpretation is closely related with Baxter's geometric solution  \cite{baxter2} differing only in the fact  that the angles involved  characterize the embedding of the lattice and not the rapidity lines themselves, which need not to be embedded in the plane as straight lines.

The reverse statement is also true, given an isoradial embedding of a lattice one can always define a Z-invariant Ising model on that lattice using the geometric angles characterizing the embedding and an elliptic function parameterization \cite{baxter2} of the coupling constants. The Ising model on isoradial embeddings considered in references \cite{mercat,kenyon3} is the restriction to the critical point of such a Z-invariant Ising model. Its critical point coupling constants $J_{ij}^c$  can be expressed in terms of the respective edge length $L_{ij}$ on the isoradial embedding of the lattice, 
\be
  \cosh{2J_{ij}^c}= \frac{2}{L_{ij}}.\label{mycrit}
\ee

This relation between coupling constants and geometry has been found previously in the study of geometric properties of regular lattices at criticality, as effective angles of the corner magnetization \cite{pearce} and shape dependent modular parameters \cite{nashoco2,me1,me2,me4}, suggesting  that the geometric interpretation  reflects an important aspect of the  Ising model critical behavior.

The paper is organized as follows: 

In Section \ref{section2} we introduce the rapidity lattice $\cal G$ and the diamond lattice $G_\di$ associated with an arbitrary planar lattice $G$. In Sections  \ref{section3}  and \ref{section4} the basic results of the Z-invariant Ising model are revisited in an embedding independent way. In  Section \ref{section5} we discuss the geometrical interpretation of the model. Finally in Section \ref{section6} our conclusions are presented.

\section{ The rapidity lattice}\label{section2}

\begin{figure}[t]
\includegraphics[width=12cm]{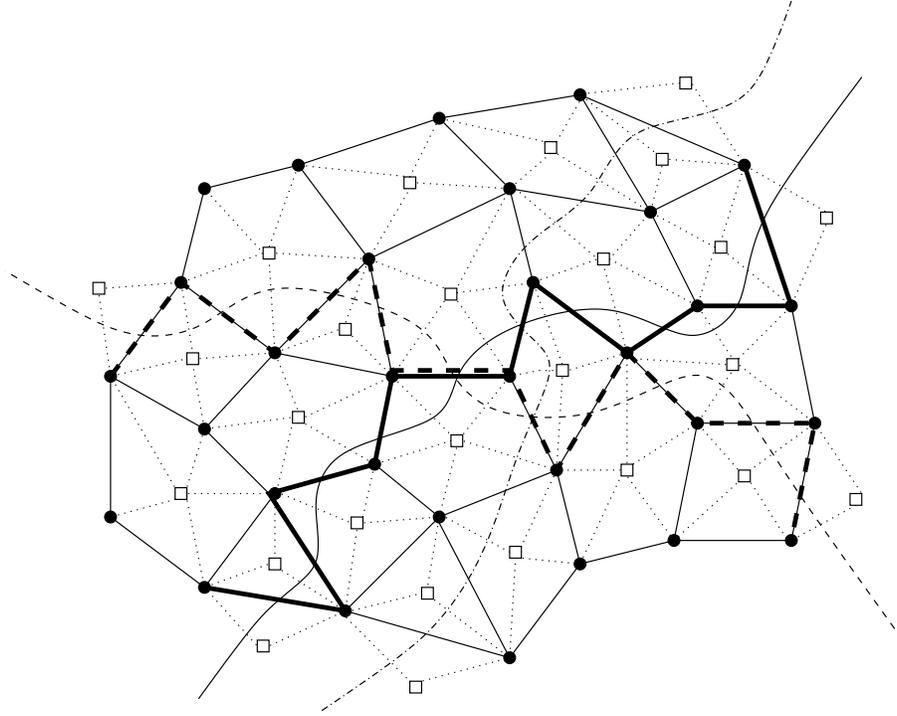}
\caption{The lattice of Fig. \ref{fig2} and the corresponding diamond graph $G_\di$ as defined on section \ref{section2}. The vertices of $G$ are shown as circles and the vertices of the dual lattice $G^*$ are shown as squares. The edges of $G_\di$ are shown in doted line. The rapidity lines of $\cal G$ are paths on the dual of $G_\di$ which are contained on a sequence of faces without turns. Three such rapidity lines are shown including the two associated with the represented zig-zag paths.}
\label{fig3}
\end{figure}

To define the rapidity lattice $\cal G$ associated with a given planar lattice $G$ it is convenient to  first introduce the diamond lattice of $G$, denoted by  $G_\di$. The diamond lattice\cite{mercat} has as vertices the vertices of $G$ and the vertices of the dual lattice $G^*$ and has edges connecting each vertex of $G$ to the vertices of $G^*$ corresponding to its neighboring faces,  an example is shown in Fig. \ref{fig3}. The faces of $G_\di$ are quadrilateral and surround an edge of $G$ and the corresponding dual edge in $G^*$.

A zig-zag path on the edges of $G$, turning alternately maximally left and right, corresponds to a path of alternating diagonals on a sequence of quadrilateral faces of  $G_\di$ in which each face connects with its two neighboring faces by two opposite edges, see Fig. \ref{fig3}. To each zig-zag path we associate a rapidity line, defined as the path in the dual of $G_\di$ which is contained by the corresponding sequence of faces. In the example of Fig. \ref{fig3} three such rapidity lines are shown. Each edge of $G$ has two  associated rapidity lines in one to one correspondence with the two zig-zag paths which intersect at that edge. 

The rapidity lattice  $\cal G$ associated with the planar lattice $G$ is formed by all the rapidity lines defined in this way. In general $\cal G$  is not  embedded in the plane as a set of straight lines.
 The calligraphic type is a reminder that we are not interested in $\cal G$  as lattice or as an embedding of a lattice in the plane but as a set of intersecting curves each with an associated free parameter, the respective rapidity. In section \ref{section5} a geometrical interpretation of this set will be given, we will see that under mild restrictions the intersection pattern of $\cal G$ together with its rapidities is a complete description of an isoradial embedding of  $G$ in the plane. An isoradial embedding of $G$, in which all its vertices are at unit length from the neighboring vertices of $G^*$, corresponds to a rhombic embedding of $G_\di$ in which all the lattice faces are rhombi of unit edge length.

\section{Z-invariance}\label{section3}

Given an arbitrary planar lattice $G$, the associated rapidity lattice $\cal G$ can always be constructed using the procedure of the previous section. However a Z-invariant Ising model can be defined on  $G$ only if the rapidity lines satisfy certain additional topological properties. In this section we review the  Z-invariant Ising model in an embedding independent way suitable for the discussion of these properties.

\begin{figure}[t]
\includegraphics[width=12cm]{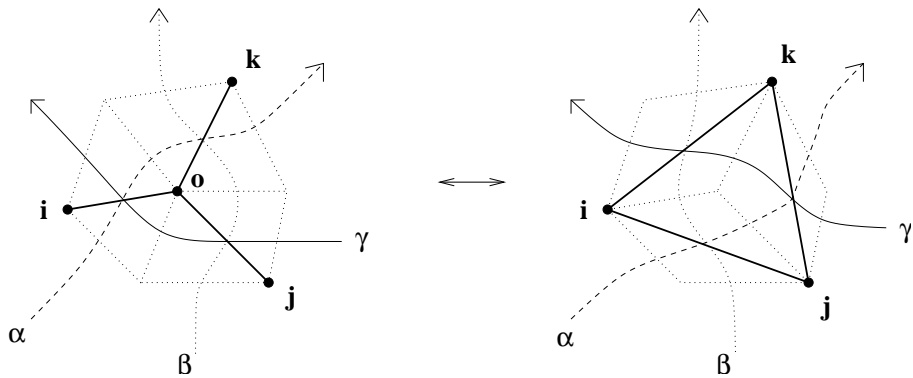}
\caption{The star-triangle transformation on the edges of $G$. The  quadrilateral faces of $G_\di$ and the three rapidity lines involved are also shown.}
\label{fig4}
\end{figure}

The Ising model on an arbitrary planar lattice  is defined in the usual way, by assigning a spin $\sigma_i=\pm 1$   to each vertex  $i$ and  a coupling constant $J_{ij}$ to the edge connecting the vertices $i$ and $j$. The model partition function is defined as
\be
    Z= \sum_{[\sigma]} \exp{\left(\sum_{<ij>} J_{ij}\, \sigma_i \sigma_j \right)}.
\ee
The model is said to be Z-invariant \cite{baxter2} if a  star-triangle deformation of the lattice,  as shown in Fig. \ref{fig4},  changes the partition function  only by a multiplicative constant \mbox{$Z_{\text{star}}= R\; Z_{\text{triangle}}$}  with $R$ being determined by the local coupling constants. 

Z-invariance places stringent conditions on the model coupling constants which can however  be solved \cite{baxter3} by introducing a parameterization in terms of  elliptic functions of modulus $k$ with $0<k<1$. Following reference \cite{baxter2} the coupling constants are  parameterized in terms of an edge dependent argument   $J_{ij}=J(\tau_{ij})$ accordingly with table \ref{table1}, by choosing a regime and a modulus $k$. The corresponding Ising model is Z-invariant if the edge arguments $\tau_{ij}$ satisfy the following conditions throughout the lattice
\bs
    &&\textit{ the sum of the edge arguments around a vertex is $2\lambda$}, \label{cond} \\
    && \textit{ the sum of the edge arguments in the boundary of a m-sided face is $(m-2)\lambda $} \nn
\es
with the multiplicative constant being, for the vertex labeling of Fig. \ref{fig4},
\be
    R=\left( \frac{2\Omega^2}{ \sinh{2J_{ij}}\sinh{2J_{jk}}\sinh{2J_{ki}}} \right)^\frac{1}{2}.
\ee 

For a straight line rapidity lattice it has been shown\cite{baxter2} that the conditions (\ref{cond}) are satisfied for edge arguments determined by the difference of the rapidities on the two straight lines crossing each edge. This result is expected not to depend on the particular embedding of the rapidity lines in the plane but only on their topological properties. We will show that the straight line requirement can be relaxed  and that the only condition we need to impose on the rapidity lines is that they do not intersect each other more than once or self-intersect. 

\begin{figure}[t]
\includegraphics[width=12cm]{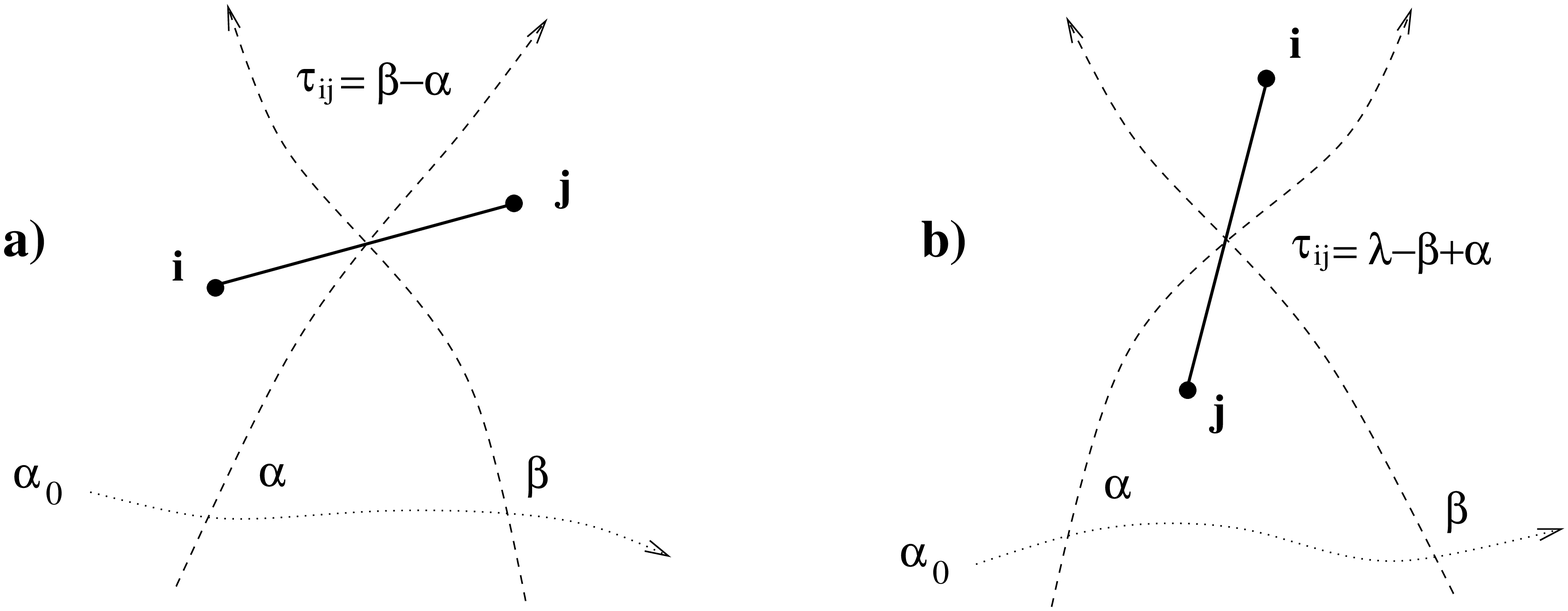}
\caption{The two possible relative orientations of an edge of $G$ and its two rapidity lines and the corresponding edge arguments. The dotted line is the reference rapidity line.}
\label{fig5}
\end{figure}

Using oriented rapidity lines \cite{perk} the argument  $\tau_{ij}$ at the edge $[ij]$ is chosen, in an embedding independent way, according to the  orientation of the edge relatively to the two rapidity lines crossing it. For un-oriented edges there are two possible orientations, shown in Fig. \ref{fig5}, and the corresponding edge arguments are chosen to be
\be
       \tau_{ij}= \begin{cases}
                     \beta-\alpha & \text{ for orientation a)}\\ 
                     \lambda+\alpha-\beta & \text{ for orientation b)}\\
                  \end{cases},   \label{angles}
\ee
where $\alpha$ and $\beta$ are the rapidities of the two rapidity lines involved.  

This choice of edge arguments satisfies the conditions (\ref{cond}) everywhere on the lattice provided that the rapidity lines do not intersect each other more than once or self-intersect. This result can be argued as follows.

 As in reference \cite{baxter1} we assume that there a rapidity line $\alpha_0$ which is crossed by all other rapidity lines  $\alpha_i$ for $i>0$.  The rapidity lines are oriented is such a way that if we look along the orientation of  $\alpha_0$  the rapidity lines $\alpha_i$  cross $\alpha_0$  from right to left.  We assume also that the lattice $G$ is formed by the intersection of the rapidity lines $\alpha_i$ after they cross the reference line $\alpha_0$. These assumptions do not place any restriction on  the lattice $G$.

\begin{figure}[t]
\includegraphics[width=12cm]{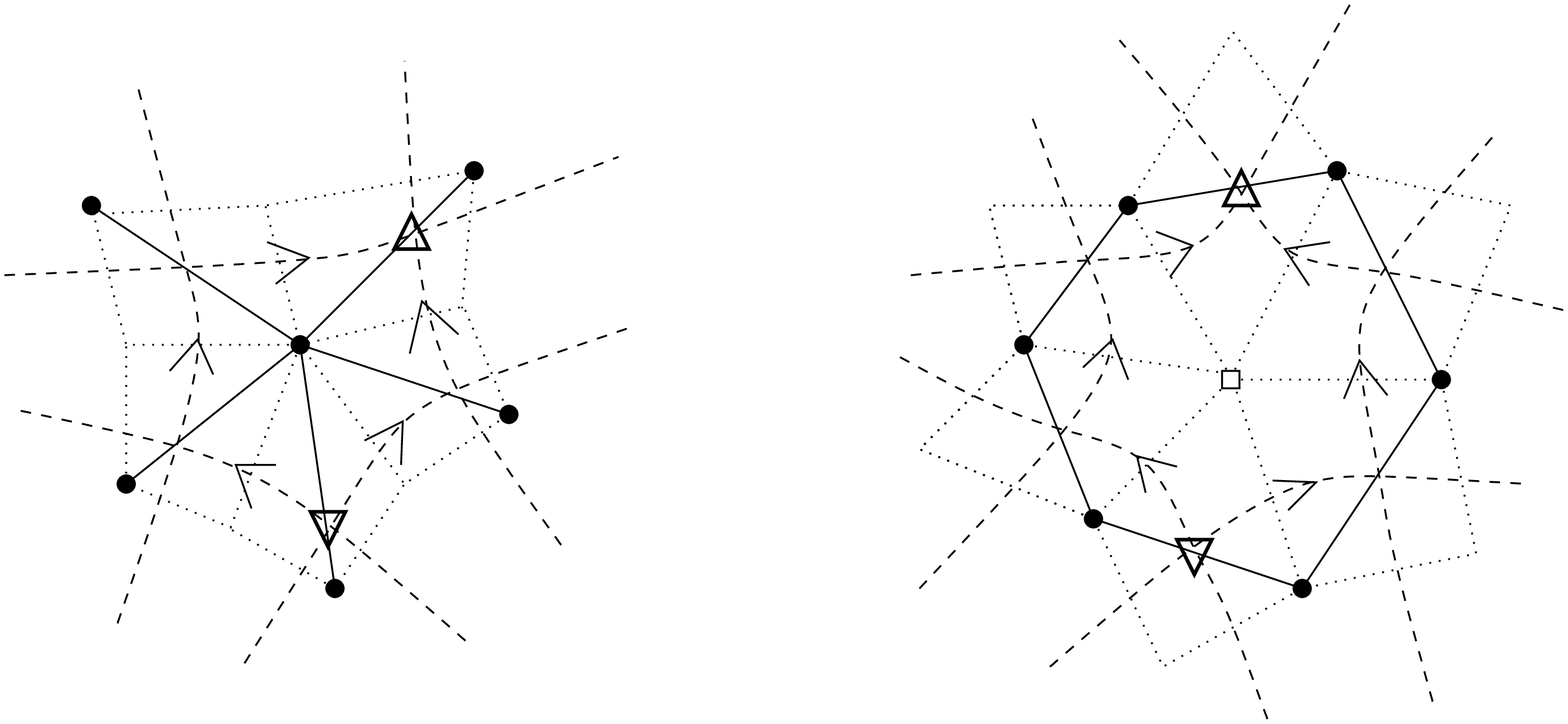}
\caption{Standard orientations of  the rapidity lines on the faces of $\cal G$ surrounding vertices of $G$ and $G^*$, shown using the notation of Fig. \ref{fig3}. The boundary edges are oriented away from a  vertex in the boundary and towards a second one, denoted by the downward and upward triangles respectively.}
\label{fig6}
\end{figure}

Consider a face of  $\cal G$ surrounding a vertex of either $G$ or $G^*$ and let its boundary edges be oriented as the respective rapidity lines. For our choice  of rapidity lines  orientations  it follows that all the boundary edges  are oriented  away from one of the boundary vertices and towards a second one, represented in the examples of Fig. \ref{fig6} by the downward and upward triangle respectively. We will call such an orientation of the boundary edges standard.

\begin{figure}[t]
\includegraphics[width=10cm]{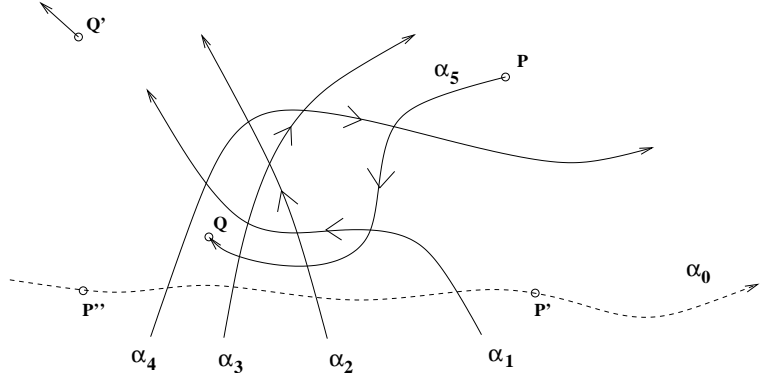}
\caption{A face of $\cal G$ with a non-standard orientation of the boundary edges. The rapidity curve $\alpha_5$ cannot be completed by joining points $P$ with either $P'$ or $P''$ and $Q$ with $Q'$ without multiple intersections or self-intersection, which by assumption are not allowed.}
\label{fig7}
\end{figure}

 Non-standard orientations are not  allowed by the restriction that the rapidity lines do not intersect each other more than once or self-intersect. Consider for instance a  face of $\cal G$ formed by the intersection of five rapidity lines, with the boundary edges oriented as shown in Fig. \ref{fig7}. Each rapidity line crosses the reference line $\alpha_0$, intersects two or more other rapidity lines  and moves towards infinity or to the boundary of a compact region enclosing the lattice $G$. Given the first rapidity line $\alpha_1$, subsequent rapidity lines will have to intersect  $\alpha_0$ to the left of $\alpha_1$ to avoid double-intersections and self-intersections. This can be done for all lines except the last one, $\alpha_5$, which should cross $\alpha_0$ at some point $P'$ or $P''$ and then reach some distant point  $Q'$. Neither of these two requirements can be met without $\alpha_5$ crossing more than once $\alpha_1$ or $\alpha_4$ or self-intersecting and this non-standard orientation is not allowed. A similar argument can be made against orientations with more than one boundary vertex  for which the boundary edges converge or diverge.

The reader can verify that for the standard orientations of $\cal G$ the edge arguments defined by (\ref{angles}) satisfy the conditions (\ref{cond}) everywhere in the lattice and the resulting Ising model is Z-invariant.  If the rapidity lines, or equivalently the lattice zig-zag paths, are allowed to intersect each other more than once or self-intersect, then the conditions (\ref{cond}) will not in general be satisfied and a Z-invariant Ising model cannot be defined.

\section{Thermodynamic properties}\label{section4}

Using  the standard arguments of Z-invariance  \cite{baxter1,baxter2} the partition function,  for real positive coupling constants, can be evaluated in the thermodynamic limit of the lattice $G$. For completeness we reproduce here these results. The partition function is given by \cite{baxter2}
\be
    Z=(4\Omega)^{\frac{1}{4}N} \exp{\left(\sum_{<ij>} \phi(\tau_{ij}) \right)} 
\ee 
where $N$ is the number of vertices on the lattice and 
\be
  \phi(\tau)= \frac{1}{2}c\tau +\int_0^\tau r(x)\left( \frac{x}{2K}+ \frac{K'}{\pi}  \frac{q'(x)}{q(x)}\right)dx
\ee
with the various quantities being defined in table \ref{table1} for the three distinct regimes.

The local magnetization can also be evaluated \cite{baxter2} and is found to be, for any spin on the lattice,
\be
    <\sigma_i>=\begin{cases}
               (1-\Omega^{-2})^{1/8} & \text{ for $\Omega^2>1$}\\
                0                    & \text{ for $\Omega^2\leq 1$}\\
             \end{cases}.
\ee 
The regime I corresponds therefore to a low temperature ordered phase, while regimes II and III correspond to high temperature disordered phases. A phase transition occurs between regime I and II for $\Omega=1$.

\section{Geometric interpretation at criticality}\label{section5}

The intersection properties that the rapidity lines, or equivalently the lattice zig-zag paths, need to satisfy in order to define a Z-invariant Ising model  on a given planar lattice are equivalent \cite{kenyon2} to the existence of isoradial embeddings of that lattice. In this section we will establish this relation between the Z-invariant model and isoradial embeddings by  considering a  geometric realization of the model.

At criticality the elliptic modulus $k$ is zero,  the complete elliptic integral $\lambda$ becomes $\pi/2$ and the coupling constants, see table \ref{table1}, are parameterized by elementary trigonometric functions
\be
       \sinh{2J_{ij}}=\tan{\tau_{ij}}.  \label{mercatf}
\ee

For non-vanishing ferromagnetic coupling constants with  edge arguments in the interval \mbox{$0\!<\!\tau_{ij}\!<\!\pi/2$}  the conditions  (\ref{cond}) can be written in terms of geometric half-angles  $\tau_{ij}=\theta_{ij}/2$ as
\bs
    \sum \theta_{ij} =2\pi &&\text{ around each vertex of $G$}  \label{conda} \\
    \sum (\pi-\theta_{ij}) = 2\pi&& \text{ around each face of $G$} \nn
\es
where the sums are respectively over all edges which meet at a given vertex of $G$ and over all edges in the boundary of a given face of  $G$.

\begin{figure}[t]
\includegraphics[width=14cm]{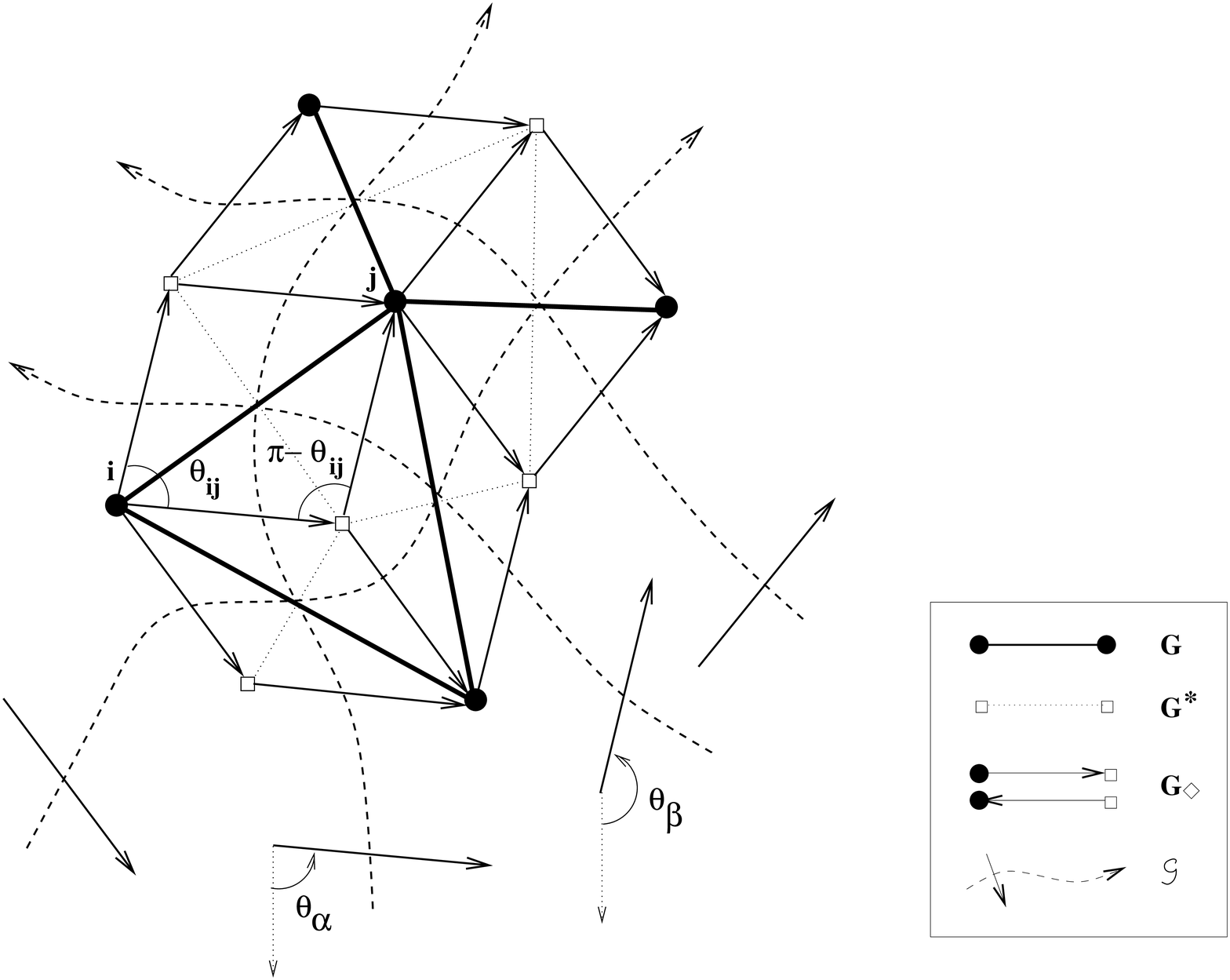}
\caption{An isoradial embedding of a lattice $G$ represented together with its dual $G^*$, the diamond lattice $G_\di$ and the rapidity lattice $\cal G$. Each edge of $G$ has an associated rhombic angle. The angle associated to each rapidity line characterizes the common direction of all edges of $G_\di$ intersected  by it.}
\label{fig8}
\end{figure}

A set of angles satisfying these conditions characterizes a rhombic embedding of $G_\di$, in  which  its quadrilateral faces are drawn in the plane as rhombi with unit edge length. The angle $\theta_{ij}$ is the inner angle of the rhombic face which surrounds the edge $[ij]$, see Fig. \ref{fig8} for an example. The equations (\ref{conda}) guarantee that the rhombi can be joined together, at the vertices of $G$ and $G^*$, in a consistent way without overlapping.  

 In such a rhombic embedding the edges of $G_\di$ crossed by a given rapidity line are  parallel and we can associated to each rapidity line an angle characterizing this direction. Any  rhombic embedding of $G_\di$  is in fact completely characterized \cite{ned} by the intersection pattern of $\cal G$, which determines the relative positions of the rhombi, and the direction of the edges crossed by each rapidity line.
The condition that the rapidity lines do not intersect each other more than once, or self-intersect, is a consistency condition needed to ensure that all interior rhombic angles are positive.

The embedding of $G$ associated with a rhombic embedding of $G_\di$ is a strictly convex isoradial embedding. Therefore if a Z-invariant Ising model can be defined on a given lattice  that lattice admits isoradial embeddings in the plane. A set of critical edge parameters in the interval $0<\tau_{ij}<\pi/2$ can always be obtained by assigning  angles to the rapidity lines in the range $]0,\pi[$ which increase relatively to the order of intersection of the rapidity lines $\alpha_i$ with the reference line $\alpha_0$.

The reverse statement is also true, if a planar lattice $G$ has an isoradial embedding in the plane then we can define a Z-invariant Ising model in $G$.  This follows from the fact that zig-zag paths on an isoradial embedding do not  intersect each other more than once or self-intersect. The corresponding Z-invariant Ising model can be defined with  edge parameters determined by the rhombic angles of the embedding  $\tau_{ij}=\lambda\theta_{ij}/\pi$.  From (\ref{mercatf}) it follows that the model coupling constants at criticality can  be  expressed in terms of the respective edges lengths according to equation (\ref{mycrit}). 
The Z-invariant Ising model with such coupling constants is similar to the geometrical model pointed out by Baxter \cite{baxter1,baxter2}, the only distinction being that  the   angles associated to the rapidity lines characterizes not the rapidity lines themselves but the parallel edges of $G_\di$ crossed by them.  
The restriction to the critical point of this Z-invariant Ising model is precisely the Ising model considered by the authors of references \cite{mercat,kenyon3} in the context of discrete holomorphy.

\section{Conclusions}\label{section6}

In this paper we considered the Z-invariant Ising model on arbitrary planar lattices introduced without reference to a straight line  rapidity lattice. We have shown that a Z-invariant model can be defined on such lattices if and only if the lattice  satisfies the topological property that zig-zag paths  do not self-intersect or intersect each other  more than once. The associated rapidity lattice is then defined with rapidity lines in one to one correspondence with  zig-zag paths.

This topological characterization has the geometric counterpart of the existence of isoradial embeddings of the lattice, establishing a close relation between the Z-invariant Ising model and the geometrical setting used in the mathematical literature to define discrete versions of analytic functions\cite{duffin,mercat,kenyon1}. 
  Our results can then be stated in a purely geometric way: a Z-invariant Ising model can be defined on an arbitrary planar lattice if and only if the lattice admits isoradial embeddings in the plane. 

The  Ising model on isoradial embeddings studied in references \cite{mercat,kenyon3} was shown to correspond to the restriction to the critical point of a particular Z-invariant Ising model. In this model, similar to Baxter's geometric solution,  each rapidity line has an associated geometric angle characterizing the direction of the edges of diamond lattice $G_\di$  which the rapidity line intersects.
The coupling constants of  model can  be expressed at criticality in terms of the inverse length  of the respective edge on the isoradial embedding (\ref{mycrit}), reproducing the relation between critical coupling constants and geometry found in studies of the corner magnetization \cite{pearce} and  shape dependent modular parameters \cite{nashoco2,me1,me2,me4} of regular lattices in non-trivial topologies.

\bigskip
\centerline{{\bf Acknowledgments}}
\bigskip

 This work was partially supported  by the EU network on ``Discrete Random Geometry'' grant HPRN-CT-1999-00161. The author profited from many useful discussions with Prof. Barry McCoy.

\begin{table}[t] \centering
\begin{tabular}{|c|ccc|}\hline
 & regime I  & regime II    & regime III \\ \hline
 $\Omega$  &  $1/k'$  &  $k'$  & $ik'/k$ \\
$\sinh{2J(\alpha)}$  & $\frac{\sn{\alpha}}{\cn{\alpha}}$ & $k'\frac{\sn{\alpha}}{\cn{\alpha}}$  & $ik'\frac{\sn{\alpha}}{\dn{\alpha}}$ \\
$\lambda$  &  $K$     &  $K$   & $K-iK'$ \\ 
$q(\alpha)$  & $\Theta_{1}(\alpha)$ & $\Theta(\alpha)$ & $\Theta(\alpha)$ \\
$r(\alpha)$ & $\frac{\dn{\alpha}}{\sn{\alpha}\cn{\alpha}}$ & $\frac{\dn{\alpha}}{\sn{\alpha}\cn{\alpha}}$ & $\frac{\cn{\alpha}}{\sn{\alpha}\dn{\alpha}}$ \\
$c$   & $\frac{k^2 K'}{\pi}$   & $-\frac{k^2 K'}{\pi}$ & $-\frac{K'}{\pi}$ \\
\hline
\end{tabular}
\caption{Parameterization of the coupling constants $J$ and related quantities in terms of elliptic functions of modulus $k$, as given in reference\cite{baxter2}. The conjugate modulus $k'=(1-k^2)^{\frac{1}{2}}$, the complete elliptic integrals $K$, $K'$ and the elliptic functions $\sn{u}$, $\cn{u}$, $\dn{u}$, $H(u)$ ,$H_1(u)$, $\Theta(u)$ and $\Theta_1(u)$  are defined in the usual way \cite{grad}.}
\label{table1}
\end{table}

\end{document}